\documentclass[10pt,a4paper]{article}
\usepackage[makeroom]{cancel}
\usepackage{bm}
\usepackage{amsmath}
\usepackage{amsfonts}
\usepackage{graphicx}
\usepackage{nicefrac}
\usepackage{authblk}
\usepackage{cite}
\usepackage[left=2.54cm,top=2.54cm,right=2.54cm,bottom=2.54cm,nohead]{geometry}
\DeclareGraphicsExtensions{.png,.jpg,.pdf}
\usepackage{epstopdf}
\usepackage{float}
\usepackage{subfigure}
\usepackage{fouriernc} 

\title{Analysis of ionic conductance of carbon nanotubes}

\usepackage{authblk}
\makeatletter
\renewcommand\AB@authnote[1]{\textsuperscript{\normalfont#1}}

\makeatother

\author[1]{P.M. Biesheuvel}
\author[2]{M.Z. Bazant}
\affil[1]{Wetsus, European Centre of Excellence for Sustainable Water Technology, Leeuwarden \& Physical Chemistry and Soft Matter, Wageningen University, The Netherlands.}
\affil[2]{Departments of Chemical Engineering and Mathematics, Massachusetts Institute of Technology, Cambridge, MA 02139, USA.}

\date{} 

\begin{document}

\renewcommand{\t}{\widetilde}
\renewcommand{\t}{}
\newcommand{\s}[1]{\mathrm{_{#1}}}

\maketitle

\begin{abstract}
We use space-charge (SC) theory (also called the capillary pore model) to describe the ionic conductance, $G$, of charged carbon nanotubes (CNTs). Based on the reversible adsorption of hydroxyl ions to CNT pore walls, we use a Langmuir isotherm for surface ionization and make calculations as function of pore size, salt concentration $c$, and pH. Using realistic values for surface site density  and pK, SC theory well describes published experimentally data on the conductance of CNTs. At extremely low salt concentration, when the electric potential becomes uniform across the pore, and surface ionization is low, we derive the scaling $G\sim \sqrt{c}$, while for realistic salt concentrations, SC theory does not lead to a simple power law for $G(c)$.

\end{abstract}

\noindent\rule{12cm}{0.4pt}

\smallskip

\medskip

\noindent The ionic conductance, $G$, of carbon nanotubes (CNTs) is of relevance for applications in membrane technology for water desalination, energy harvesting and energy conversion~\cite{Hinds, Mauter, Vilatela, Liu, Striolo, Mattia}.  Secchi \emph{et al.}~\cite{Secchi_Arxiv,Secchi_PRL}  recently reported the first experimental results for $G$ of single carbon nanotubes of different radii and lengths, in a large salt concentration range (1-1000 mM) and at several values of pH. The observed dependence of $G$ on pH, and the absence of a plateau in $G$ at low salinity, were taken as evidence that CNTs acquire a surface charge by reversible adsorption of hydroxyl ions from water. A theoretical analysis led to a 1/3$^\text{rd}$ power-law scaling of $G$ with salt concentration, which is supported by the data.

\smallskip
In the present work, to describe the same data of Secchi \emph{et al.}~\cite{Secchi_Arxiv,Secchi_PRL}, we use the general classical dilute solution theory for long and thin capillary pores, combining the extended Nernst-Planck equation with the Stokes equation for fluid flow and the Poisson-Boltzmann (PB) equation for the structure of the electrical double layer (EDL), evaluated in radial direction. This model was developed by Osterle and co-workers~\cite{Gross&Osterle,Fair&Osterle} and is known as the capillary pore model, or space charge (SC) theory. SC theory is based on ideal Boltzmann statistics of ions as point charges, and assumes validity of the equilibrium Poisson-Boltzmann (PB) equation in the radial, $r$, direction~\cite{Sasidhar&Ruckenstein,Westermann_Clark,Hawkins_Cwirko,
Wang,Peters,Catalano_JPCM,Catalano_slip}. SC theory also includes an axial salt concentration gradient, but this effect is neglected in the present analysis. Secchi \textit{et al.}~\cite{Secchi_Arxiv,Secchi_PRL} use SC-theory with several simplifications to arrive at an analytical expression for $G$ versus pore size and salt concentration. For CNTs they introduce the key idea that the surface charge depends on pH (in the external bath) and surface potential, via a model for the reversible adsorption of hydroxyl ions.

\smallskip

The structure of this report is as follows. We present the SC theory for the conductance $G$ and show model simplifications when the Donnan approach, or uniform potential (UP) model~\cite{Catalano_JPCM,Catalano_slip,Bocquet&Charlaix,Tedesco} is used, valid for highly overlapped electric double layers (EDLs). We derive a scaling law of $G$ with salt concentration in the low-salinity limit. We assess the assumptions made in the derivation of Secchi \textit{et al.}'s analytical solution. Finally we combine the full SC theory with a Langmuir isotherm for ionizable surface charge to describe data of Secchi \emph{et al.}~\cite{Secchi_Arxiv,Secchi_PRL} for the conductance of CNTs. 

\smallskip

When we neglect axial gradients in salt concentration, SC theory only requires a (numerical) solution of the PB-equation in a cylindrical nanopore, to calculate potential $\psi$ as function of $r$-coordinate, 
\begin{equation} \label{eq:PB}
\frac{1}{r}\frac{\partial}{\partial r}\left(r\frac{\partial \psi}{\partial r}\right) = \frac{1}{\lambda\s{D}^2}\sinh\psi
\end{equation}
where 
\begin{equation}
\lambda\s{D}=\sqrt{\frac{\varepsilon V\s{T}}{2 F \t{c}}}
\end{equation}
is the Debye length, $\varepsilon$ the dielectric constant, $\varepsilon=\varepsilon\s{w}\varepsilon_0$, $V\s{T}=RT/F=k\s{B}T/e$ the thermal voltage, and $c$ the salt concentration in the external baths, in mol/m$^3$. Unless otherwise noted, all parameters are dimensional (except for $\psi$ and Pe$^0$). 
Boundary conditions for Eq.~(\ref{eq:PB}) are
\begin{equation}\label{eq:BoundaryPB}
\left.\frac{\partial\psi}{\partial r}\right|_{r=0} = 0 \hspace{1.2em} , \quad \left.\frac{\partial\psi}{\partial r}\right|_{r=R} = +\frac{\sigma}{\varepsilon \> V\s{T}}
\end{equation}
where $\sigma$ is the wall charge density in C/m$^2$.
\smallskip

The ionic conductance of a nanopore, $G$ (in A/V) is the ratio of current over voltage drop, in the absence of axial gradients in concentration or pressure~\cite{Secchi_Arxiv}. In SC theory, $G$ is given by~\cite{Gross&Osterle,Sasidhar&Ruckenstein,Westermann_Clark,Wang, Peters,Balme,Catalano_slip}
\begin{equation} \label{eq:G_SC}
G=4\pi \> \mu\s{D} \> \t{c} \> F \> \ell^{-1} \cdot \left( \int^R_0 r \,\cosh\psi \, \textrm{d}r + \text{Pe}^0 \cdot \int^R_0 r \sinh\psi \> (\psi_\mathrm{w} - \psi)\> \textrm{d}r \right)
\end{equation}
where Pe$^0=\left( \vphantom{\mu\s{D}}\varepsilon V\s{T} \right) / \left( \mu\s{w}\mu\s{D} \right)$ is the ``normalization'' P{\'e}clet number \cite{Nielsen}, where $\mu\s{D}=D/V\s{T}$ 
and $D$ is the ion diffusion coefficient, assumed to be the same for both ions. 
Furthermore, $\mu\s{w}$ is the dynamic viscosity of water, $\psi\s{w}$ the dimensionless electric potential at the tube surface (wall), $F$ is Faraday's constant, and $\ell$ the length of the nanotube. Parameter settings in this report are $D=2\cdot 10^{-9}$ m$^2$/s, $\mu\s{w}=1$ mPa$\cdot$s and $\varepsilon\s{w}=78$ (Pe$^0=0.228$). Eq. (\ref{eq:G_SC}) assumes zero wall slip and equal ion diffusion coefficients. For the general case with wall slip and $D_+ \neq D_-$, see refs.~\cite{Catalano_JPCM, Catalano_slip}. In Eq. (\ref{eq:G_SC}) the first term is a ``direct Ohmic conductance'' where the conductivity is proportional to the pore-averaged ion concentration, while the second term 
accounts for the streaming current carried by charge advection, where the fluid is set in motion by the electric field (electro-osmosis). In Ref. \cite{Stein_PRL} these two terms are called the conductive and convective contributions to the current, while in Ref. \cite{Heyden_PRL} only the second, convective, term is considered.

\medskip

Secchi \textit{et al.}~\cite{Secchi_Arxiv,Secchi_PRL}, as in Ref.~\cite{Bocquet&Charlaix}, use an expression for $G$ which can be derived from Eq. (\ref{eq:G_SC}) when the second (convective, or electro-osmotic) term is neglected, and the Donnan equation 
\begin{equation}\label{eq:donnan}
\t{\sigma}=R \> \t{c} \> F \> \sinh{\psi}
\end{equation}
is used, which is an overall electroneutrality balance over the pore. Eq. (\ref{eq:donnan}) can be used when the electric double layers (EDLs) that are extending from the pore walls become sufficiently overlapped, and ${\partial \psi}/{\partial r}$, thus wall charge, $\sigma$, is not too high. In this limit, the pore potential $\psi$ becomes invariant with position in the pore and thus equal to $\psi\s{w}$. Thus, Eq. (\ref{eq:donnan}) is valid when $\psi$ varies weakly with position, valid in the high EDL overlap regime, when the Debye length $\lambda\s{D}$ is much larger than pore size $R$, and when surface charge is not too high. Combining Eqs. (\ref{eq:G_SC}) and (\ref{eq:donnan}) leads to 

\begin{equation}\label{eq:G_donnan}
G\cdot\frac{ \ell }{\pi \mu\s{D} R^2 F} =  2\>\sqrt{\left( \frac{\t{\sigma}}{FR}\right)^2+\t{c}^2} \>\>+\>\>\frac{\t{\sigma}^2 }{2F\mu\s{w}\mu\s{D}}
\end{equation}
of which Secchi \textit{et al.} only use the first term (Eq. (3) in ref.~\cite{Secchi_PRL}, similar to Eq. (38a) in Ref.~\cite{Bocquet&Charlaix}). 

\medskip

The Donnan approximation is valid at very low salt concentration (and not too high charge), when the Debye length is much larger than the pore size, and also at very high salt concentration, when the potential is close to zero at all positions in the pore. The electro-osmotic term (second term in Eq. (\ref{eq:G_SC})) can be neglected when the fluid is at rest at all radial positions, which however is generally not the case.  
Analyzing the importance of the electro-osmotic term in the full SC theory, we find that e.g. in Fig. 1B for pH 6 ($R$=14 nm pore), its contribution to the total conductance $G$ is 25\% at 1 mM salt but drops to 4\% at 1 M.

\medskip

\begin{figure}
\centering
\includegraphics[scale=0.53]{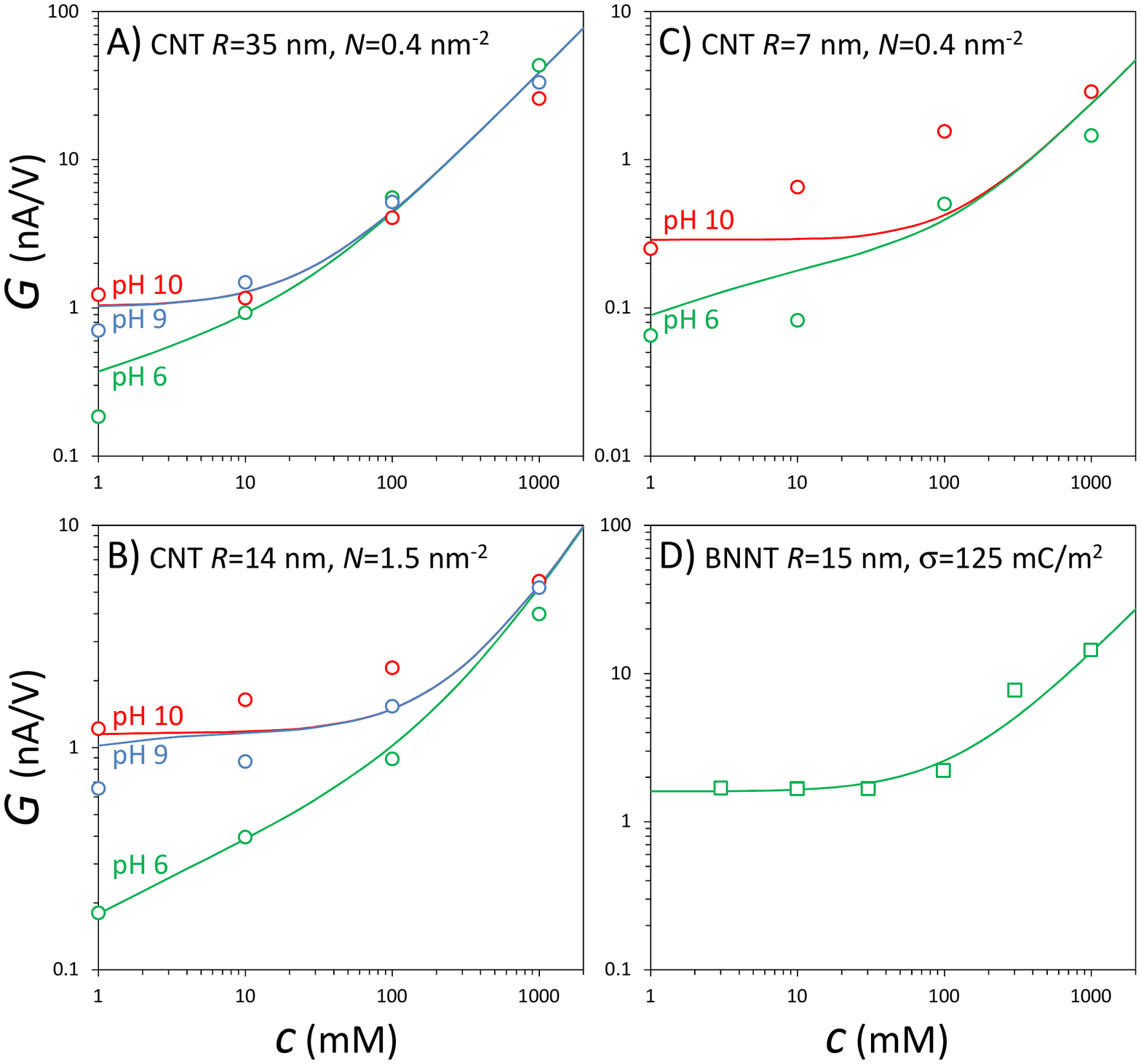}
\caption{Conductance $G$ of single carbon nanotubes (A-C), and BNNTs (D), as function of salt concentration $\t{c}$, fitted with space charge theory. For BNNT a fixed wall charge is assumed; for CNTs a Langmuir ionization isotherm ($\text{pK}\enskip 4$). Data from ref.~\cite{Secchi_Arxiv}. Tube length $\ell$: A) 1.5 B) 2.0 C) 1.0 D) 0.8 $\mu$m.}\label{fig1}
\end{figure}

For a material with a fixed wall charge, 
the above theory suffices. However, for a surface with ionizable charge, an implicit relationship between surface charge and surface potential must be included which is based on a chemical model of ionization of the surface. This is a classical approach in colloid science \cite{Biesheuvel_JPCM}, also applied to ionic flow through membranes by Koh and Anderson~\cite{Koh&Anderson} in their study of electrolyte conductance through 15-50 nm radius polyelectrolyte-adsorbed track-etched pores in 7 micron thick mica-sheets. This approach was pioneered for CNTs by Secchi \textit{et al.}~\cite{Secchi_Arxiv,Secchi_PRL}. 

\medskip

To describe ionization by a site-binding model, the Langmuir 1$-$pK adsorption isotherm is often used which considers a maximum number of ionizable sites, $N$, and includes the entropy of the distribution between charged and uncharged sites~\cite{Siria,Koopal_EA,Heyden_PRL,Biesheuvel_JPCM,Andersen_PRL,Koh&Anderson}. It is a two-parameter model based on $N$ and pK, 
and can be extended to include a Stern capacity~\cite{Catalano_slip}. For a surface that charges negatively (either by hydroxyl ion adsorption, or for instance by the ionization of carboxylic acid groups), the Langmuir isotherm is given by
\begin{equation}\label{eq:ioniz}
\t{\sigma}=-eN\cdot\frac{1}{1+10^{\text{pK}-\text{pH}_\infty}\cdot\exp\left(-\psi \s{w}\right)}
\end{equation}
where $\text{pH}_\infty$ is pH in bulk solution outside the nanotube. To use this equation {\it globally} in a theory of charged nanopores with ion transport, the surface composition (charge) must be in equilibrium with the pH in the external bulk solutions, which -- at the very least -- requires that pH is the same on both sides of the CNT, and that no axial concentration gradients of ion concentrations develop along the pore. These are indeed the typical assumptions made in the literature on ionic conductance. Instead, when concentration- and pH-gradients do develop through the nanopore, the full equations for transport for cations, anions, and proton/hydroxyl ions must be solved to find how $\sigma$ and pH change {\it locally} along the pore~\cite{Andersen_PRL}.

\smallskip

In the limit of a low ionization, $\alpha=\left|\t{\sigma}\right|/eN \ll 1$, the Langmuir model, Eq. (\ref{eq:ioniz}), can be written as~\cite{Koh&Anderson,Biesheuvel_JPCM, Secchi_Arxiv}

\begin{equation}\label{eq:low_ionization}
\t{\sigma}=\t{\sigma}_\infty\> \cdot \exp\left(\psi\s{w}\right)
\end{equation} 
where $\t{\sigma}_\infty=-e\cdot N\cdot 10^{\text{pH}_\infty-\text{pK}}$ is the charge at zero surface potential, such as attained for very high background salinity. Note, $\psi\s{w}$, $\sigma$ and $\sigma_\infty$ have a negative value for a surface that charges negatively. 
To arrive at an analytical solution, we combine Eq. (\ref{eq:low_ionization}) with an appropriate EDL-model. In the low salt-limit, this is the Donnan model that was already discussed, Eq. (\ref{eq:donnan}). For any non-zero value of $\t{\sigma}_\infty$ there is some value of $\t{c}$ below which $|\psi\s{w}|\sim |\psi|$ is large enough for $\sinh\left(\psi\s{w}\right)$ to be approximated by $\tfrac{1}{2}\exp\left(\psi\s{w}\right)$, and combination of Eqs. (\ref{eq:donnan}) and (\ref{eq:low_ionization}) then results in
\begin{equation} \label{eq:sigma}
|\t{\sigma}|=\sqrt{ \tfrac{1}{2} F R \> | \t{\sigma}_\infty |  \>}\cdot \sqrt{\t{c}}
\end{equation}
which is the counterion-only limit, or ``good co-ion exclusion limit''~\cite{Balme}. Eq.~(\ref{eq:sigma}) shows that in this limit the surface charge becomes smaller when salt concentration goes down.  

\smallskip

At very low $\t{c}$, and with $|\t{\sigma}| \propto \sqrt{\t{c}\>}$ according to Eq. (\ref{eq:sigma}), in Eq. (\ref{eq:G_donnan}) only the first term within the square root remains as a contribution to $G$. Making use of Eq. (\ref{eq:sigma}) we thus arrive at a square-root scaling relation,
\begin{equation}\label{eq:G_SC_ii}
G = \alpha \cdot \sqrt{\t{c}\>},
\end{equation}
where  
$\alpha=\tfrac{1}{4}\>\pi\sqrt{2} \> \mu\s{D} \> R^{3/2} \ell^{-1}{F}^{1/2} \sqrt{|\t{\sigma}_\infty|}$. 
Though mathematically interesting, we emphasize that this scaling is not attained under practical conditions. For instance, for the theory line in Fig. 1B at pH 6, the power-law slope, $s$, is $s=0.34$ at $\t{c}=1$ mM, $s=0.43$ at \mbox{$\t{c}=1$ $\mu$M}, and $s=0.49$ at $\t{c}=1$ nM. Therefore, the limiting square-root scaling is only reached in extremely dilute solutions, where the continuum hypothesis would also break down within the CNT.

\smallskip

Instead, Secchi \textit{et al.} arrived at a 1/3$^\text{rd}$ power law scaling, which matches their analytical model from a very low to a quite high salt concentration (approx. 100 mM), in line with the experimental data, so let us consider how this result was derived. 
The Supplementary Information of Secchi \textit{et al.} explains that in the derivation use is made of the Gouy-Chapman (GC) equation, which describes the structure of a planar isolated EDL, which is given by
\begin{equation}\label{eq:GC}
\t{\sigma}=\sqrt{8\varepsilon R\s{g}T\t{c}\>} \> \sinh\left(\tfrac{1}{2}\psi\s{w}\right).
\end{equation}
%
The GC model can be combined with Eq. (\ref{eq:low_ionization}) to show that for any non-zero $\t{\sigma}_\infty$, below some value of $\t{c}$, $|\psi\s{w}|$ will be high enough that the sinh-function can be replaced by $\tfrac{1}{2}\times$ the exp-function, after which combination of Eqs. (\ref{eq:low_ionization}) and (\ref{eq:GC}) results in
\def\yT{\vphantom{\left. T \right)}}
\begin{equation}\label{eq:bocquet_1}
\t{\sigma}=\left(2\varepsilon R\s{g}T\right)^{1/3}\> {\t{\sigma}_\infty \yT}^{1/3}\>{c\yT}^{1/3}
\end{equation}
which shows a 1/3$^\text{rd}$ order scaling of $\sigma$ with salt concentration $c$. Note that Eq. (\ref{eq:bocquet_1}) is derived using the GC model for {\it thin} double layers on a planar surface, valid for $\lambda\s{D}<<R$. 
Next, to obtain an expression for conductance $G$,  Eq. (\ref{eq:bocquet_1}) is combined with only the first term in Eq. (\ref{eq:G_donnan}), which is valid for {\it thick} double layers $\lambda\s{D}>>R$, neglecting the electro-osmotic contribution. Because $\sigma$ scales with $c^{1/3}$, at sufficiently low $c$ the term $c^2$ in Eq. (\ref{eq:G_donnan}) can be neglected, so $G$ is proportional to $\sigma$ given by Eq. (\ref{eq:bocquet_1}) and thus $G$ scales with $c^{1/3}$, as derived by Secchi et al.~\cite{Secchi_Arxiv,Secchi_PRL}.

\smallskip

To analyze their data, Secchi \textit{et al.} introduce a prefactor $C\s{0}$ which encompasses all right-hand terms in Eq. (\ref{eq:bocquet_1}) except for $c^{1/3}$, and thus, as identified by Secchi \textit{et al.}, must scale with pH according to $C\s{0}\propto 10^{\text{pH}/3}$, while otherwise it must be constant, independent of CNT radius. The set of values for $C\s{0}$  derived from fitting Eqs. (\ref{eq:G_donnan}) and (\ref{eq:bocquet_1}) to each data set separately, are presented in Fig. 1 in Suppl. Inf. of Secchi \textit{et al.}. Here we see that the data for 3.5 nm tubes are in line with this pH-scaling, but this is not the case for other data sets (for instance, the data for 14 nm tubes have a scaling in $C\s{0}$ versus pH not with $1/3\sim0.31$ but rather with $\sim 0.12$). Furthermore, at each pH-value, $C\s{0}$ has a quite large variation in the derived values (obtained for tubes of different diameter), up to a factor of 25 difference between the highest and lowest value in $C\s{0}$ at pH 10. Though there is not a definite trend, $C\s{0}$ more or less decays with increasing pore size, whereas it should be pore size-independent. Using Eq. (\ref{eq:bocquet_1}) and the definition of $C\s{0}$ given by $\Sigma \lambda\s{B}^2=C\s{0} \left(\rho\s{s}\lambda\s{B}^3\right)^{1/3}$ (where $\Sigma$ is the surface charge in m$^{-2}$, $\lambda\s{B}$ the Bjerrum length for which we use $\lambda\s{B}=$0.72 nm, and $\rho\s{s}$ the salt concentration in m$^{-3}$) we can convert the measured value of $C\s{0}$ to the corresponding maximum charge density $|\sigma_\infty|$ (at high salinity and the same pH). For 14 nm CNTs at pH 6 (panel B) we arrive with $C\s{0}\sim 2.7$ at $|\sigma_\infty|\sim 40$ C/m$^2$, or equivalently, at >200 fixed charges per nm$^2$, which is clearly an unrealistically high number. 

\medskip

As we show below, when we solve the full SC theory with the full Langmuir equation and compare with the data for CNTs, we obtain a reasonably good fit to most of the data sets without fitting a separate value of $C\s{0}$ to each data set, but using as sole adjustable parameters the pK value (for which we use pK 4 throughout, similar to pK for carboxylic acid groups) and the maximum site density of charged groups (for which we use either $N=0.4$ or $N=1.5$ nm$^{-2}$). 

As shown in Fig.~1, which is similar to Fig. 1 in ref.~\cite{Secchi_Arxiv}, the quality of the model fit varies from moderate to good. For panels A and C we used the lower value for the site density, $N=0.4$ nm$^{-2}$ and a higher value in panel B ($N=1.5$ nm$^{-2}$). Both values are realistic (for instance, silica has a significantly higher density of ionizable groups of $N \sim 8$ nm$^{-2}$). The value of $N=1.5$ nm$^{-2}$ recalculates to a maximum surface charge (at high pH and high salinity) of $-240$ mC/m$^2$ but dependent on pH and salt concentration, the actual surface charge density is much lower, for instance for the calculation in Fig. 1B, for pH 6, charge varies from $-39$ mC/m$^2$ at 1 mM, to $-80$, $-145$, and $-209$ mC/m$^2$ at 10, 100 and 1000 mM. 

\medskip

In ref.~\cite{Secchi_PRL}, Fig. 1 includes additional data for pores with radii of $R=3.5$ and 10 nm. Here we reproduce this figure as Fig. 2 (panels A and B are the same as in Fig. 1 above) and use for the data in the new panels C) and D) a site density of $N=1.5$ nm$^{-2}$. For CNT with a radius of $R=10$ nm (pH 4), the fit is perfect, see Fig. 2C. However, comparison of SC-theory to data for CNTs with a radius of $R=3.5$ nm is not adequate at pH 8 and pH 10, see Fig. 2D. In contrast to the other data sets, conductance $G$ does not yet converge to a single curve at salt concentrations beyond 1 M, as SC-theory would predict. Clearly, in the experiments with CNTs of $R=3.5$ nm at pH 8 and pH 10, there is an additional effect which is not included in the present formulation of SC-theory, such as perhaps a non-negligible fluid wall slip in CNTs~\cite{Secchi_Nature_2016}. Also, for such thin CNTs it becomes likely that axial gradients develop in pH and salt concentration along the pore, just as for pores in a nanofiltration membrane. i.e., the CNT works as a desalination device for which the full two-dimensional version of SC theory must be solved~\cite{Gross&Osterle,Fair&Osterle,Peters}.

\begin{figure}
\centering
\includegraphics[scale=0.53]{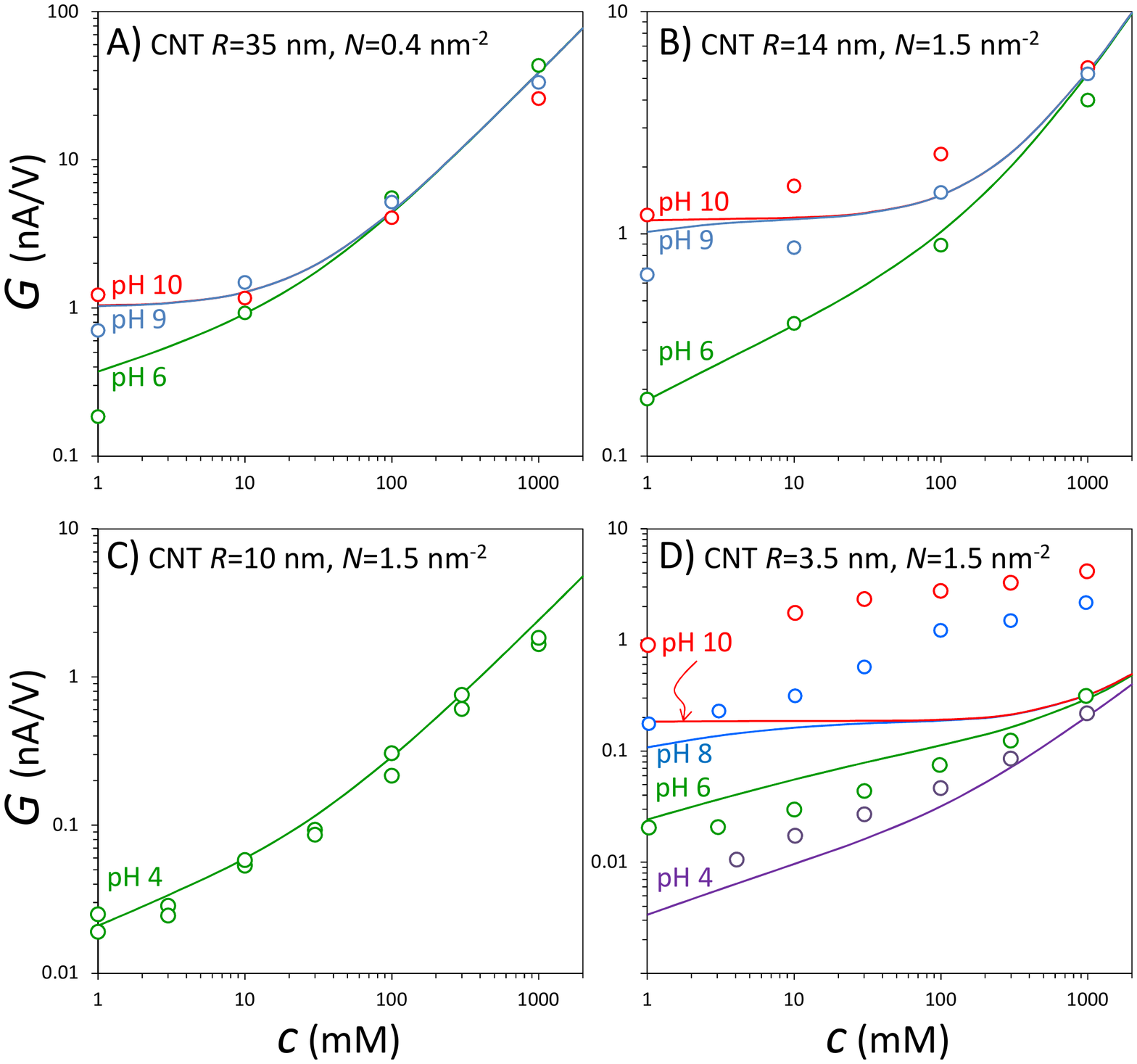}
\caption{Conductance $G$ of single carbon nanotubes as function of salt concentration $\t{c}$, fitted with space charge theory and a Langmuir ionization isotherm ($\text{pK}\enskip 4$). Data from ref.~\cite{Secchi_PRL}. Tube length $\ell$: A) 1.5 B) 2.0 C) 2.0 D) 3 $\mu$m. Panels A and B are the same as in Fig. 1.}\label{fig2}
\end{figure}

\smallskip

Finally we analyze data by Secchi \textit{et al.} on the conductance $G$ of single BNNTs, where we assume a fixed wall charge density $\t{\sigma}$. Here we find that the data for conductance $G$ versus salt concentration $\t{c}$ in Fig.~\ref{fig1}D can be accurately described by the full SC theory with a wall charge of \mbox{$\t{\sigma}=125$ mC/m$^2$} in line with a value of \mbox{$\t{\sigma}=100$ mC/m$^2$} given by Siria \textit{et al}.~\cite{Siria} (pH 5). 

\smallskip

In conclusion, classical space-charge theory can be a useful theoretical tool to describe ionic conductance of charged (carbon) nanotubes. In combination with a Langmuir adsorption isotherm for OH$^-$-adsorption, data for the conductance of single carbon nanotubes are reasonably well described, using realistic, constant parameter settings for pK and surface site density, across a range of different salt concentrations and nanotube geometries.

\end{document}